\documentclass[12pt]{article}

\usepackage{amsmath,amssymb,amsfonts}

\usepackage{times}
\usepackage{bm,mathrsfs,dsfont,color}
\usepackage{natbib}
\usepackage{graphicx, float}
\graphicspath{{plots/}}
\usepackage[font=small]{caption}

\usepackage[hyphens]{url} 
\usepackage[colorlinks=false, hidelinks]{hyperref}

\usepackage{authblk}

\usepackage{multirow}

\def\T{{ \mathrm{\scriptscriptstyle T} }}
\newcommand{\ppg}{f_{\mbox{\textsc{pg}}}}

\begin{document}

%% Here are the title, author names and addresses
\makeatletter
\renewcommand\section{\@startsection {section}{1}{\z@}%
                                   {-3.5ex \@plus -1ex \@minus -.2ex}%
                                   {2.3ex \@plus.2ex}%
                                   {\centering\normalfont\large\scshape}}
                                   
\renewcommand\subsection{\@startsection {subsection}{1}{\z@}%
                                   {-3.5ex \@plus -1ex \@minus -.2ex}%
                                   {2.3ex \@plus.2ex}%
                                   {\centering \normalfont \scshape}}                                   
\makeatother

\title{Efficient posterior sampling for Bayesian Poisson regression}

\author[1]{Laura D'Angelo\thanks{ laura.dangelo@unimib.it }}
\author[2]{Antonio Canale}

\affil[1]{Department of Economics, Management and Statistics; University of Milano-Bicocca}
\affil[2]{Department of Statistical Sciences; University of Padova}

\date{}

\maketitle

\abstract{
Poisson log-linear models are ubiquitous in many applications, and one of the most popular approaches for parametric count regression. In the Bayesian context, however, there are no sufficient specific computational tools for efficient sampling from the posterior distribution of parameters, and standard algorithms, such as random walk Metropolis-Hastings or Hamiltonian Monte Carlo algorithms, are typically used. 
Herein, we developed an efficient Metropolis-Hastings algorithm and importance sampler to simulate from the posterior distribution of the parameters of Poisson log-linear models under conditional Gaussian priors with superior performance with respect to the state-of-the-art alternatives. 
The key for both algorithms is the introduction of a proposal density based on a Gaussian approximation of the posterior distribution of parameters. Specifically, our result leverages the negative binomial approximation of the Poisson likelihood and the successful P\'olya-gamma data augmentation scheme. Via simulation, we obtained that the time per independent sample of the proposed samplers is competitive with that obtained using the successful Hamiltonian Monte Carlo sampling, with the Metropolis-Hastings showing superior performance in all scenarios considered.

{\center \textbf{Keywords: }}
Count data; Horseshoe prior; Importance sampling; Log-linear models; Metropolis-Hastings; P\'olya-gamma. \vfill

\section{Introduction}
\label{sec:intro}

Poisson log-linear models are common in statistics and represent one of the most popular choices to model how the distribution of count data varies with predictors. A typical assumption is that, under an independent sample of counts, $y_1, \dots, y_n$, the probability mass function of the generic $y_i$ conditionally on a $p$-dimensional vector of covariates $x_i$ is
\begin{equation}
f(y_i \mid \lambda_i) =  \frac {\lambda_i^{y_i}}{{y_i}!}e^{-\lambda_i}, \quad \log(\lambda_i) = x_i^T \beta,
\label{eq:model0}
\end{equation}
where $\beta$ is a $p$-dimensional vector of unknown coefficients. Linking the linear predictor $x_i^T \beta$ and the parameter $\lambda_i$ with the logarithm represents the most natural choice, as the logarithm is the canonical link for the Poisson family~\citep{nelder1972glm}.
This model has broad application in several fields, including medicine and epidemiology~\citep{Frome1983, frome1985, Hutchinson2005}, manufacturing process control~\citep{lambert1992}, analysis of accident rates~\citep{Sarath1990, Miaou1994}, and crowd counting~\citep{chan2009}, among others.

In the Bayesian context, model \eqref{eq:model0} does not enjoy any conjugacy property and, thus, regardless of the prior used, the posterior distribution of $\beta$ is not available in close form. Consequently, inference is conducted using Markov Chain Monte Carlo (MCMC) methods, which obtain a sample from the posterior distribution of the parameters. Several approaches have focused on how to easily obtain the posterior distribution of the coefficients of Poisson models without requiring complex tuning strategies or long computation times. 
In the context of count-valued time series,~\cite{fruhwirth2006} proposed a formulation of the model based on two levels of data augmentation, to derive an efficient approximate Gibbs sampler. \cite{fruhwirth2009} exploited a data augmentation strategy to simplify the computation of hierarchical models for count and multinomial data.
Data augmentation strategies have also been employed in the case of models for multivariate dependent count data~\citep{karlis2005, bradley2018}. 
However, the simplest Poisson regression in \eqref{eq:model0} still lacks a specific and efficient algorithm to sample from the posterior distribution of the parameters $\beta$ for any prior choice, making the Metropolis-Hastings \citep{hastings1970monte} or Hamiltonian Monte Carlo (HMC) \citep{neal2011mcmc} algorithms the only available options.

On the other hand, several efficient computational strategies for binary regression models have been proposed in the literature. Using the probit link, \cite{albert_chib_1993} proposed an efficient data augmentation based on a latent Gaussian variable, while the more recent contribution by \cite{polson_scott_2013} exploited the canonical logit link, introducing an efficient P\'olya-gamma data augmentation scheme.
Leveraging \cite{polson_scott_2013} approach, we propose a novel approximation of the posterior distribution that can be exploited as proposal distribution of a Metropolis-Hastings algorithm or as importance density of an importance sampling for Poisson log-linear models with conditional Gaussian prior distributions on the regression parameters. With conditional Gaussian prior, we refer to a possibly hierarchical prior with conditional distribution $\beta \sim \mathrm{N}(b, B)$, with $b$ and/or $B$ random. Examples include straightforward Gaussian prior distributions with informative $(b,B)$ fixed using prior information, and scale mixtures of Gaussian where $b$ is set to zero and the variance has a suitable hierarchical representation, such as the Bayesian lasso prior \citep{park2008bayesian}, the horseshoe prior, and its extensions \citep{carvalho2010horseshoe,Piironen2017}.

More specifically, we introduce an approximation of the posterior density that exploits the negative binomial convergence to the Poisson distribution. Thanks to this result, we are able to leverage the P\'olya-gamma data augmentation scheme of~\cite{polson_scott_2013} to derive an efficient sampling scheme. 
In the next section, we introduce and discuss the proposed algorithms, starting from the definition of an \emph{approximate} posterior distribution whose sampling can be performed straightforwardly.
Sampling from this \emph{approximate} posterior is then used as proposal density for the Metropolis-Hastings or importance sampler. The performances of the proposed algorithms in terms of computational efficiency is compared with that of state-of-the-art methods in a simulation study. The paper concludes with two illustrative applications.

\section{Efficient posterior sampling strategies}
\label{sec:2}

\subsection{Approximate posterior distribution}\label{sec:2.1}

Assume $y_1,\dots,y_n$ is an independent sample of counts from model ~(\ref{eq:model0}). 
We introduce an approximation of the posterior density which exploits the negative binomial convergence to the Poisson distribution, i.e., we approximate the $i$-th contribution to the likelihood function $f(y_i \mid \lambda_i)$ with $\tilde{f}_{r_i}(y_i \mid \lambda_i)$ where 
\begin{equation}
\tilde{f}_{r_i}(y_i \mid \lambda_i) = 
\binom{r_i + y_i -1}{r_i-1} \left( \frac{r_i}{r_i + \lambda_i}\right)^{r_i} \left(\frac{\lambda_i}{r_i+ \lambda_i}\right)^{y_i},
\label{eq:approx1}
\end{equation}
which corresponds to the probability mass function of a negative binomial random variable with parameter $r_i$, the number of failures until the experiment is stopped, and success probability $\lambda_i/(r_i+\lambda_i)$. As $r_i$ approaches infinity, this quantity converges to a Poisson likelihood.

Following \cite{polson_scott_2013}, we implement a data augmentation scheme based on the introduction of $n$ P\'olya-gamma random variables. A random variable $X$ is said to follow a P\'olya-gamma distribution with parameters $\xi>0$ and $\zeta \in \mathbb{R}$, in the following denoted as $X\sim \mbox{PG}(\xi,\zeta)$, if 
\begin{equation*}
	X \overset{D}{=} \frac{1}{2\pi^2} \sum_{k=1}^{\infty} \frac{g_k}{(k-1/2)^2 + \zeta^2/(4\pi^2)},
	\end{equation*}
where the $g_k \sim \mathrm{Gamma}(\xi,1)$ are independent gamma random variables, and where $\overset{D}{=}$ indicates equality in distribution. A key identity is that the binomial likelihood, parameterized by log-odds, can be represented as a scale mixture of Gaussians with respect to a P\'olya-gamma distribution, i.e.
\begin{equation*}
	\frac{(e^z)^{\tau}}{(1+e^z)^{\xi}} = 2^{-\xi} e^{z(\tau - \xi/2)} \int_0^{+\infty} e^{-\omega z^2 / 2} \ppg(\omega; \xi, 0)\, d\omega,
	\end{equation*}
where $\ppg(\cdot; \xi,\zeta)$ denotes the density of a P\'olya-gamma with parameters $( \xi,\zeta)$.

This is also true for the negative binomial likelihood parameterized as in \eqref{eq:approx1}: indeed, we can rewrite each $i$-th contribution to the approximate likelihood \eqref{eq:approx1} by introducing augmented P\'olya-gamma random variables $\omega_i \sim \mbox{PG}(y_i+r_i,0)$, i.e.,
\begin{equation}
\begin{aligned}
\tilde{f}_{r_i}(y_i \mid \beta) = & \binom{r_i+y_i-1}{r_i+y_i} 2^{-(y_i+r_i)} \exp\left\{ \frac{(x_i^\T \beta - \log r_i)(y_i-r_i)}{2} \right\} \times \\ 
& \int_{0}^{+\infty} \exp\left\{-\omega_i\frac{(x_i^\T \beta - \log r_i)^2}{2} \right\} \ppg(\omega_i; y_i + r_i,0) \, d\omega_i.  
\end{aligned}\label{eq:bineg_DA}
\end{equation}
More details on the derivation of this expression can be found in Appendix A of the Supplementary Material.

In what follows, we assume that prior knowledge about the unknown $\beta$ parameters is represented by a conditionally Gaussian prior, i.e. $\beta \sim \mathrm{N}(b, B)$, with a possible hierarchical representation for the parameters $b$ and $B$. Examples include default informative Gaussian with fixed $(b,B)$ or scale mixtures of Gaussian where $b$ is set to zero and the variance has a suitable hierarchical representation \citep{park2008bayesian, carvalho2010horseshoe,Piironen2017}.

The \emph{approximate} posterior based on the conditionally Gaussian prior $\beta \sim \mathrm{N}(b, B)$ and approximate likelihood $\prod_{i=1}^n \tilde{f}_{r_i}(y_i \mid \beta)$ is consistent with the successful Gibbs sampler of \cite{polson_scott_2013}; i.e., sampling from the \emph{approximate} posterior is equivalent to sampling iteratively from the following full conditionals 
\begin{equation}
\omega_i|\beta \sim \mbox{PG}(y_i + r_i, x_i^\T \beta - \log r_i), \qquad
\beta | y,\omega \sim \mathrm{N}_p(m_{\omega}, V_{\omega}),
\label{eq:approx0}
\end{equation}
where $V_{\omega} = (X^\T\Omega X + B^{-1})$ and $m_{\omega} = V_{\omega}(X^\T \kappa + B^{-1}b)$, with $\Omega = \mathrm{diag} \{\omega_1, \dots, \omega_n\}$ and $\kappa = (\omega_1 \log r_1 + (y_1-r_1)/2, \dots, \omega_n \log r_n + (y_n-r_n)/2)$. 
Derivation of this data augmentation strategy can be obtained in a straightforward manner starting from the model formulation in Eq. \eqref{eq:bineg_DA} and following the original strategy of \cite{polson_scott_2013}.

The adherence of this approximate posterior to the true posterior highly depends on the values of $r_i$, with larger values of $r_i$ resulting in better approximations. However, when employing this result in posterior sampling, large values of $r_i$ imply longer computation time due to the computational cost of sampling P\'olya-gamma random variables with large parameters. Although the specific choice of $r_i$ remains an open point---discussed later in Section \ref{sec:tuning_param}---in the context of MCMC sampling, we propose to reduce the computational burden related to the sampling of $n$  P\'olya-gamma  random variables marginalizing the Gaussian distribution in \eqref{eq:approx0} with respect to the related P\'olya-gamma density conditioned on $\beta^{(t-1)}$,  the last available $\beta$ sampled. Since this marginalization is not in a closed form we introduce a second level of approximation of the true posterior. Specifically, we introduce $q(\beta \mid \beta^{(t-1)})$ a density that depends on $\beta^{(t-1)}$, defined as the first-order Taylor expansion of the marginalized Gaussian distribution, i.e. 
\begin{equation}
\begin{aligned}
q(\beta\mid \beta^{(t-1)}) & = 
(2\pi)^{-p/2} \mathrm{det}(V_{\mathrm{E}(\omega)})^{-1/2} \exp\left\{-\frac{1}{2}(\beta - m_{\mathrm{E}(\omega)})^\T V_{\mathrm{E}(\omega)}^{-1} (\beta - m_{\mathrm{E}(\omega)})\right\},
\end{aligned}
\label{eq:approx2}
\end{equation}
where $V_{\mathrm{E}(\omega)} = (X^\T \tilde{\Omega} X + B^{-1})$, $m_{\mathrm{E}(\omega)} = V_{\mathrm{E}(\omega)}(X^\T \tilde{\kappa} + B^{-1}b)$, 
$\tilde{\Omega} = \mathrm{diag} \{ \mathrm{E}(\omega_1),\dots,\mathrm{E}(\omega_n)\}$, $\tilde{\kappa} = (\mathrm{E}(\omega_1) \log r_i + (y_1-r_i)/2, \dots, \mathrm{E}(\omega_n) \log r_i + (y_n-r_i)/2)$, and for each $i = 1, \dots, n$ the conditional expectation of each $\omega_i$ is simply 
\begin{equation*}
\mathrm{E}\left(\omega_i\right) = \frac{r_i+y_i}{2(x_i^\T\beta^{(t-1)}-\log r_i)}\left(\frac{e^{x_i^\T\beta^{(t-1)}}-r_i}{e^{x_i^\T\beta^{(t-1)}}+r_i}\right),
\end{equation*}
or equivalently
\begin{equation}
\beta\mid \beta^{(t-1)} \sim \mathrm{N}(m_{\mathrm{E}(\omega)}, V_{\mathrm{E}(\omega)}).
\label{eq:proposal_distr}
\end{equation}
The above construction is eventually used as the building block of efficient Metropolis-Hastings and importance sampling algorithms, as described in the following sections.

\subsection{Metropolis-Hastings sampler}
\label{sec:mcmc}

We employ the above sampling mechanism as the proposal density in a Metropolis-Hastings algorithm. Consistent with this, at each iteration of the MCMC sampler, an additional step that accepts or rejects the proposed draw is introduced. Specifically, we assume that conditionally on the current state of the chain $\beta^{(t-1)}$, a new value $\beta^*$ is sampled from \eqref{eq:proposal_distr}. Then, the acceptance probability 
\begin{equation}
\alpha(\beta^*,\beta^{(t-1)}) = \min\left\{ 1, \frac{\pi(\beta^*\mid y)}{\pi(\beta^{(t-1)}\mid y)}\frac{q(\beta^{(t-1)}\mid \beta^*)}{q(\beta^{*}\mid \beta^{(t)})} \right\},
\label{eq:acceptanceMH}
\end{equation}
is evaluated to decide whether to accept or reject the proposed $\beta^*$, where $\pi(\beta^*\mid y)$ is the exact posterior distribution of $\beta$ given the sample $y_1,\dots,y_n$.

To compute the acceptance probability in \eqref{eq:acceptanceMH}, the forward and backward transition densities $q(\beta^*\mid \beta^{(t-1)})$ and $q(\beta^{(t-1)}\mid \beta^*)$ must be computed. Consistent with this,  approximation~(\ref{eq:approx2}) is particularly useful: without it, it would be necessary to compute the marginal density where the P\'olya-gamma random variables are integrated out. However, the marginalization with respect to the  P\'olya-gamma density does not lead to a closed form expression; thus, the Metropolis-Hastings algorithm cannot be defined. 

Clearly, for increasing $r_i$ the proposal density \eqref{eq:proposal_distr} is closer to the true full conditional distribution; hence, the related acceptance rate will be higher, and the Metropolis-Hastings algorithm will be similar to a Gibbs sampler. 
On the other hand, setting this parameter to get a lower acceptance rate can result in smaller autocorrelation, and hence a better mixing \citep{robert2010}. We discuss an approach to choose $r_i$ balancing these two extremes in Section~\ref{sec:tuning_param}.

\subsection{Adaptive importance sampler}
\label{sec:is}

The sampling mechanism \eqref{eq:proposal_distr} can also be exploited within the context of importance sampling, where the posterior expectation of a function of the parameter $\beta$, $\mathrm{E}\left( h(\beta) \right) = \int h(\beta)\, \pi(\beta \mid y) d\beta$ is evaluated via Monte Carlo integration without direct sampling from $\pi(\beta \mid y)$. To this end, the general approach is to define an importance density $q(\beta)$ that is used to sample values $\beta^{(1)}, \dots, \beta^{(T)}$, which are eventually averaged to obtain an approximation of $\mathrm{E}\left( h(\beta) \right)$ through
\[
\widehat{\mathrm{E}( h(\beta))} = \frac{1}{T} \sum_{t=1}^T \tilde{w}(\beta^{(t)}) h(\beta^{(t)}),
\]
with weights 
\begin{equation*}
\tilde{w}(\beta^{(t)}) = \frac{\pi(\beta^{(t)} \mid y)}{q(\beta^{(t)})}.
\end{equation*}

The efficiency of this algorithm is determined by the ability of the importance density to sample values relevant to the target distribution.
To improve this aspect, we modify the original algorithm and, instead of using a fixed density $q$, at each iteration we consider an adaptive proposal. Specifically, we make use of \eqref{eq:approx2} as proposal density, but, unlike the Metropolis-Hastings algorithm,  we update it only when the last sampled value moves towards a region with a higher posterior probability. Denoting with $\beta^{c}$ the conditioning value, at each iteration we sample a new value from $q(\beta^{(t)}\mid\beta^c)$, and if $\pi(\beta^{(t)}\mid y)>\pi(\beta^{c}\mid y)$, we set $\beta^{c} = \beta^{(t)}$. Thus, the importance density is adaptively updated and the weights become
\begin{equation*}
	\tilde{w}(\beta^{(t)}) = \frac{\pi(\beta^{(t)} \mid y)}{q(\beta^{(t)} \mid \beta^{c})}.
	\end{equation*}
Notice that if one already had an estimate of the mode of the posterior (e.g. resulting from a variational approach, as in \cite{arridge2018}, or a numerical optimization), such a value could be used as a fixed conditioning value. This approach would lead to the advantage of avoiding evaluation and update of $\beta^c$ at each iteration; however, finding the mode is not trivial in general, and the overall feasibility and convenience should be evaluated on a case-by-case basis.

\subsection{Tuning parameters $r_i$}
\label{sec:tuning_param}

The values of the parameters $r_i$, $i=1,\dots,n$, have to be tuned to balance the trade-off between acceptance rate and autocorrelation in the Metropolis-Hastings, and to control the mixing of the weights in the importance sampler. However, tuning $n$ parameters is not practical, especially for moderate to large $n$.
The first simple solution sets all  parameters equal to a single value $r$, however, in our experience, this resulted in a low effective sample size for some of the sampled chains.

As an alternative strategy, we choose to tune instead the distance of the proposal density from the target posterior.
As the expression of the posterior distribution is unknown, we control the distance between the Poisson and negative binomial likelihood.  
Based on \citet{Teerapabolarn2012}, we consider the upper bound of the relative error between the Poisson and negative binomial cumulative distribution functions. This result is particularly useful owing to its simplicity, which allows to analytically derive adequate parameters to bound the error to a specific value. Specifically, if $Y$ is a Poisson random variable with mean $\lambda_i$ and $V$ is a negative binomial random variable with parameters $r_i$ and $p_i$, as defined in Section~\ref{sec:2.1}, we have the following result:
\begin{equation}
\mathrm{sup}_{y_i\geq 0} \left\lvert \frac{\mbox{Pr}(Y\leq y_i)}{\mbox{Pr}(V\leq y_i)} -1 \right\rvert = e^{-\lambda_i} p_i^{-r_i} - 1.
\label{eq:relerr}
\end{equation}

Hence, by setting an upper bound $d$ for the distance between the Poisson and negative binomial distribution, all the values of the parameters $r_i$ can be automatically derived to obtain a proposal density whose distance from the target posterior is constant for every $y_i$, even for heterogeneous data. Under our notation $p_i=\lambda_i/(r_i+\lambda_i)$, thus
$d = e^{-\lambda_i} ( 1 + r_i/\lambda_i)^{r_i} - 1,$
which is solved by
\begin{equation}
r_i = - \lambda \log c \cdot
\left\{ \log c + \lambda \cdot\mathrm{W}\left( \frac{-c^{-1/\lambda} \log c}{\lambda} \right) \right\}^{-1},
\label{eq:r_solution}
\end{equation}
where $c=e^{\lambda}(d^2+1)$ and $\mathrm{W}(\cdot)$ is the Lambert-W function \citep{W}, which can be computed numerically using standard libraries. 
Hence, in the algorithm, at the beginning of each iteration, the values $r_1, \dots, r_n$ are computed according to~(\ref{eq:r_solution}) conditionally on the current value of $\beta$. %Then, 

The relative error in \eqref{eq:relerr}  may appear a quite unusual choice to measure the ``closeness'' between two distributions, given the existence of more formal distances such as the Kullblack-Leibler divergence or the Wasserstein distance.
However, our choice is driven by the need to have a simple  analytic expression that allowed us to invert the relationship, i.e., to fix the discrepancy and derive the parameter $r_i$ that ensures it. Notably, the relationship between the relative error and, for example, the Kullblack-Leibler divergence is strictly monotone for all values of $\lambda_i$.

\section{Numerical illustrations}
\label{sec:illustr}

\subsection{Synthetic data}
\label{subsec:sim}

We conducted a simulation study under various settings to compare the efficiency of the proposed Metropolis-Hastings and importance sampler with that of state-of-the-art methods. 
For comparison, we focused on the Hamiltonian Monte Carlo approach, as implemented in the Stan software \citep{stan}. Adopting, for example, the successful Metropolis-Hastings with a standard random walk proposal would indeed require the tuning of $p$ parameters, which becomes almost unfeasible for moderate to elevate $p$. 
The proposed methods are implemented via the R package \texttt{bpr}, %\citep{bpr}, 
which is written in efficient C++ language exploiting the \texttt{Rcpp} package \citep{RCPP} and available from the Comprehensive R Archive Network \citep{bpr} and in the repository
at \url{github.com/laura-dangelo/bpr}.

Data were generated from a Poisson log-linear model with sample sizes $n \in \{ 25, 50, 100,$ $200\}$ and number of covariates $p \in \{ 5, 10, 20\}$. 
Specifically, for each combination of $n$ and $p$, we consider 50 independent $n$ dimensional vectors of counts where each $y_i$ ($i = 1\dots, n$) is sampled from a Poisson distribution with mean $\lambda_i = e^{x_i^\T \beta}$, with common parameter $\beta$. The covariates were generated from continuous or discrete/categorical random variables under the constraints that the continuous variables have mean zero and variance one and that $1\leq \lambda_i \leq 200$. Reproducible scripts to generate the synthetic data are available at \url{github.com/laura-dangelo/bpr} and as Supplementary Materials.

Two prior distributions for the coefficients $\beta$ were assumed, namely a vanilla Gaussian prior with independent components $\beta_j\sim \mathrm{N} (0,2)$, $j=1,\dots,p$, and the more complex horseshoe prior~\citep{carvalho2010horseshoe} which allows for the following  conditionally Gaussian representation
\begin{gather*}
\beta_j \mid \eta^2_j, \tau^2 \sim \mathrm{N}(0,\eta^2\tau^2) \\
\eta \sim \mathrm{C}^+(0,1), \quad \tau \sim \mathrm{C}^+(0,1),
\end{gather*}
for $j=1,\dots,p$, where $\mathrm{C}^+(0,1)$ is the standard half-Cauchy distribution. To implement the samplers under the horseshoe prior, we used the details of~\cite{makalic2016horseshoesampler}, and fixed the global shrinkage parameter $\tau$ to the ``optimal value'' $\tau_n(p_n) = (p_n/n)\sqrt{\log(n/p_n)}$, where $p_n$ is the number of non-zero parameters~\citep{vanderpas2017}. 

Each method introduced in Section \ref{sec:2}  was run for 10000 iterations with  5000 of them discarded as burn-in. The convergence of each algorithm was assessed by graphical inspection of the trace plots of the resulting chains. The convergence was satisfactory for all simulations and comparable for all algorithms, as no systematic bias was found in the posterior mean of the estimated parameters.

To assess the efficiency of the proposed methods, we used a proxy of the time per independent sample, which is estimated as the total time (in seconds) necessary to simulate the entire chain, over the effective sample size of the resulting chain. For the proposed adaptive importance sampler, an estimate of the effective sample size was obtained using the quantity $\sum_{t=1}^T w(\beta^{(t)})^2 / (\sum_{t=1}^T w(\beta^{(t)}))^2$, which takes values between 1 and $n$~\citep{robert2010}. 
Notably, the burn-in samples were removed from the chains before computing the effective sample size. Thus, the obtained times per independent sample  do not represent exactly the number of seconds necessary to generate one independent sample---they rather represent an overestimate. Nonetheless, this approach provides a robust and fair comparison between the different competing algorithms. The experiment has been run on a macOS machine with 32 GB DDR4 2400 MHz RAM, CPU Intel Core i7 4.2 GHz, running R 4.1.1.

\begin{figure}
	\begin{center}
		\includegraphics[width = \textwidth]{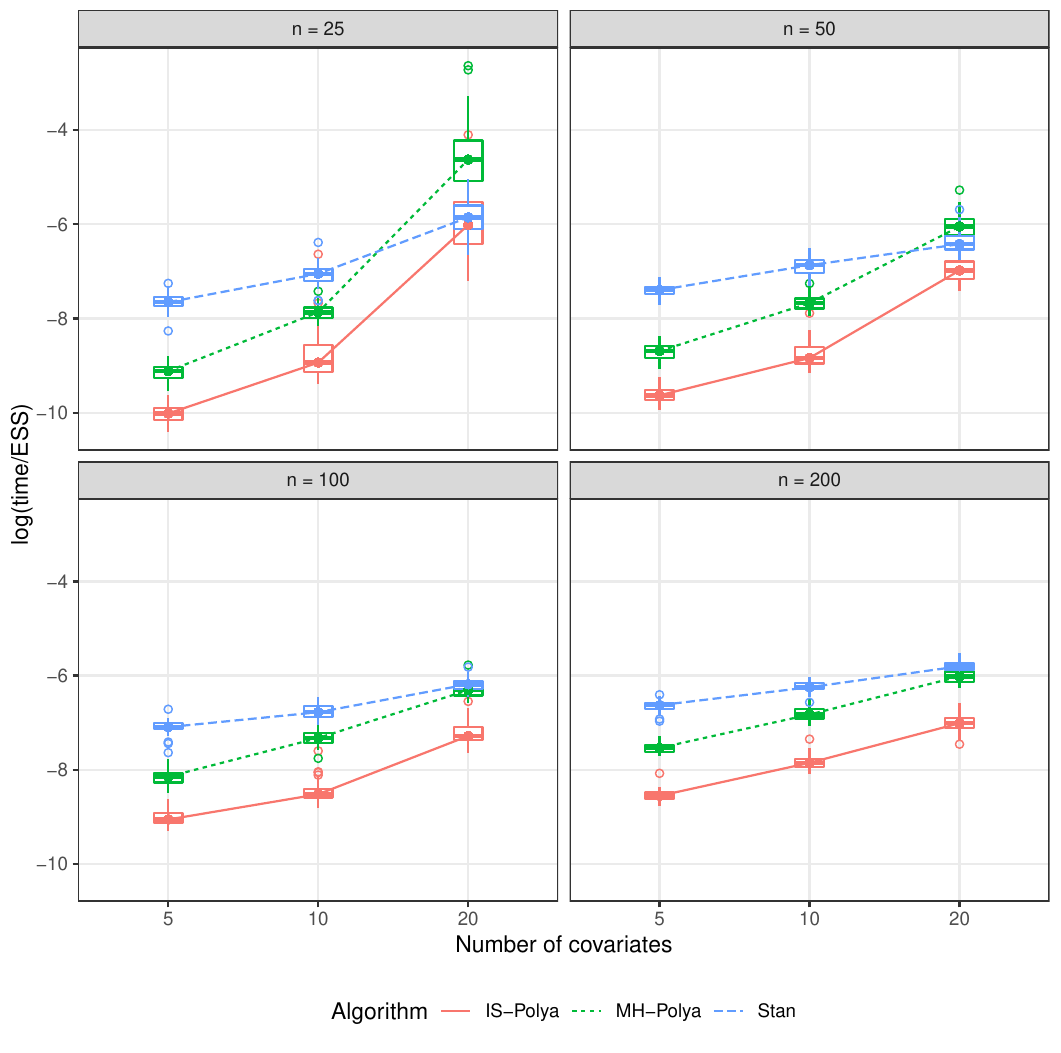}
		\caption{Time per independent sample (in logarithmic scale) for the three algorithms. For each combination of $n$ and $p$ the boxplots represent the distribution of the (log) time (in seconds) over the effective sample size using a Gaussian prior, over 50 replications. \label{fig:time_ess}}
	\end{center}
\end{figure}

Figure~\ref{fig:time_ess} and \ref{fig:time_ess_horseshoe} show, for each combination of $n$ and $p$, the distribution of the median time per independent sample for the three algorithms computed on the 50 replications under a Gaussian and horseshoe priors, respectively. The plots are presented in the logarithmic scale for clarity.

For the Gaussian prior the performances of the proposed algorithms are better than those obtained using the HMC implemented in Stan, for small values of the dimension $p$. For $p=20$, instead, the performances of the HMC are quite competitive with respect to the importance sampling and broadly comparable to the proposed efficient Metropolis-Hastings algorithm. Notably, the differences are less evident with increasing sample size.

\begin{figure}
	\begin{center}
		\includegraphics[width = \textwidth]{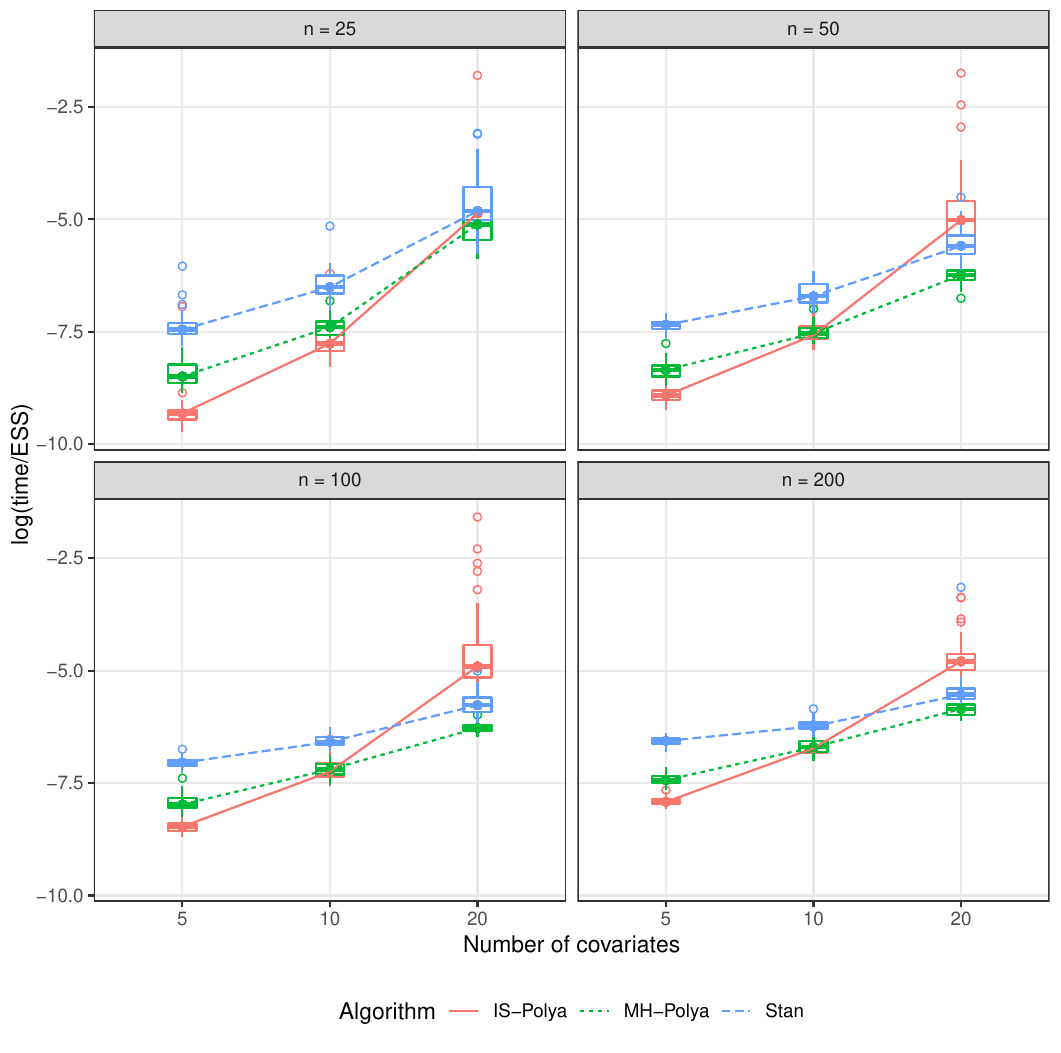}
		\caption{Time per independent sample (in logarithmic scale) for the three algorithms. For each combination of $n$ and $p$ the boxplots represent the distribution of the (log) time (in seconds) over the effective sample size using the horseshoe prior, over 50 replications.
			\label{fig:time_ess_horseshoe}}
	\end{center}
\end{figure}
For the horseshoe prior, the proposed Metropolis-Hastings presents a stable superior performance with respect to the HMC sampler implemented in Stan for each sample size $n$ ad number of covariates $p$. The performance of the importance sampler remains competitive. As previously observed for the Gaussian prior, the differences are less evident for increasing sample size.

Our experience suggests that for increasing dimension, the performance of the methods gets worse but remains competitive for moderate $p$ beyond the range considered here. Results for $p=50$ are available in Appendix B of the Supplementary Material.

\subsection{Spike train data}

Herein, we illustrate the proposed sampling method on spike train data, which describe the neurons' activity in response to stimulation.
This type of data is relatively new, and it arises from the observation of brain activity through the technique of calcium imaging. Thanks to this technique, it is possible to investigate the association between a set of external conditions and the neurons' response.
The data set was generated using a small subset of the the Allen Brain Observatory~\citep{allen}, which is a survey of the activity of neurons in mice in response to visual stimulation. Specifically, we applied the deconvolution method of \citet{jewell2019} as described in~\citet{vries2020} to count the activations of each neuron, to analyze how they are affected by the experimental conditions and other covariates available from the study.

%In the original data set, for each neuron the fluorescent calcium traces are recorded, which is a proxy of the neuronal activity, under different experimental conditions. From these traces, it is of interest to detect and analyze the activations of neurons, which correspond to transient spikes of the intracellular calcium level. We applied the method reported by~\citet{jewell2019} as described in~\citet{vries2020} to extract and count the activations of each neuron, to understand how they are affected by the experimental conditions and the location of the neurons in the brain.

In the context of neural studies, this approach is usually referred to as ``encoding models'', as the interest is to predict the activity of a population of neurons in response to a given stimulus and other experimental conditions. \citet{paninski2007} reviewed some methods commonly employed in encoding problems. Generalized linear models, which are flexible and yet interpretable, are one of the fundamental tools for investigating the response of neurons to external factors. In particular, the authors assert that assuming a Poisson distribution is a plausible assumption to model spike counts; hence, we regressed the estimated number of neurons’ activation on several continuous and categorical covariates available from the study.

\begin{figure}
	\centering
		\includegraphics[width = \linewidth]{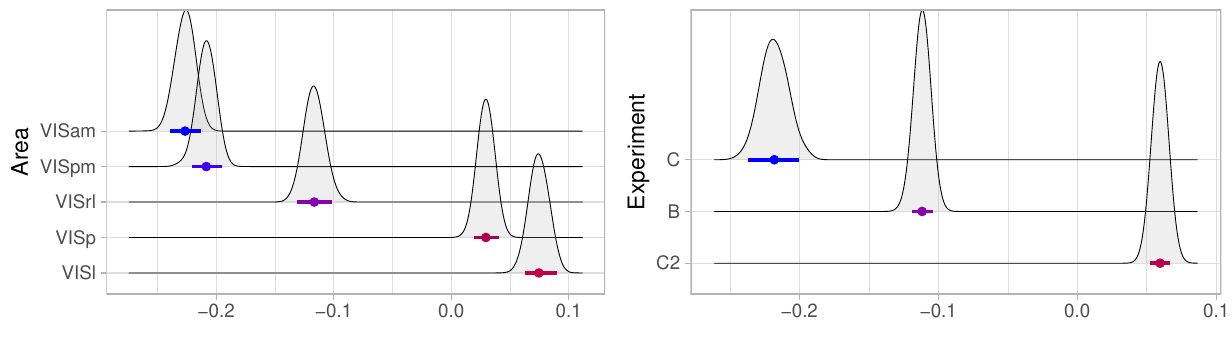}
		\includegraphics[width = \linewidth]{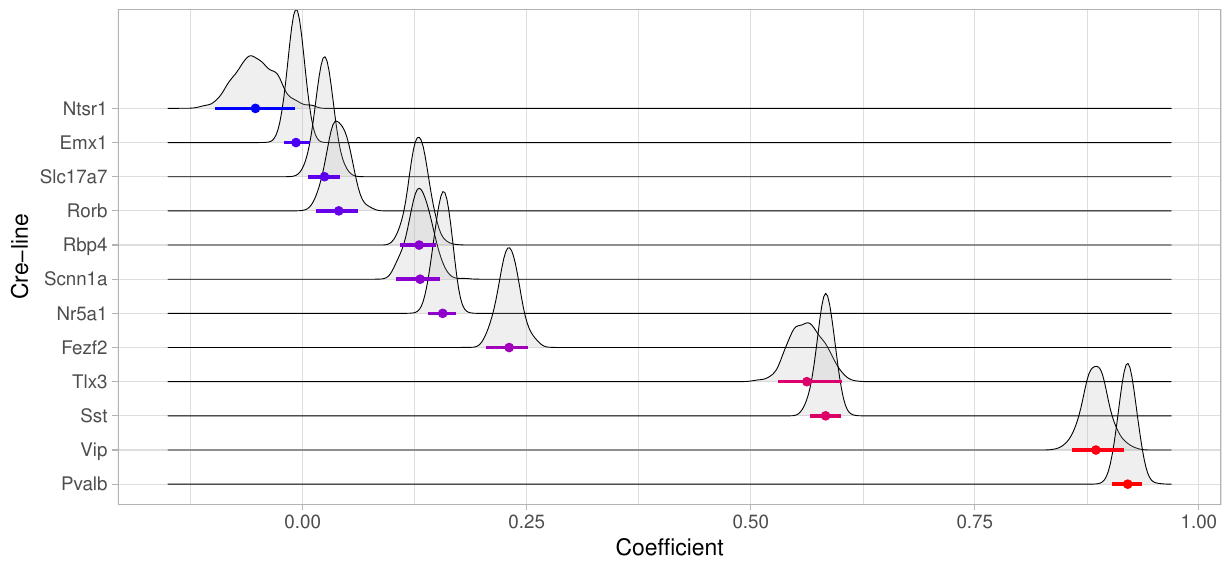}   
	\caption{Coefficients of the regression on the calcium imaging data set: posterior density, with the posterior mean and 95\% credible interval (colored dot and segment).} \label{fig:calcium_coeff}
\end{figure}

The covariates are the depth of the neuron, the area of the visual cortex where the neuron is located (factor with 6 levels), the cre transgenic mouse line (factor with 13 levels), and the type of visual stimulation (factor with 4 levels). 
The depth of the neurons is discretized to 22 levels, ranging from 175 to 625 microns, thus, we could obtain a data set having a full factorial design with 5 replications for each available covariate combination. Moreover, we included a quadratic term of the depth to improve the fitting. The obtained data set is made of 920 observations on 23 variables.

We ran the proposed Metropolis-Hastings algorithm for 9000 iterations, discarding the first 5000 as burn-in. The computation time was 98 seconds.
We employed the Metropolis-Hastings sampler rather than the importance sampler as it was more stable. Our experience suggests that this is a general behavior and thus we recommend using the Metropolis-Hastings algorithm for high $p$. 
The posterior estimates of the coefficients of the dummies on three categorical variables are shown in Figure~\ref{fig:calcium_coeff}; and for the numeric covariate depth, the posterior mean and 95\% credible intervals were equal to $-2.72\times 10^{-3}$ $(-2.90\times 10^{-3}, -2.42\times 10^{-3})$ for the linear term, and $5.59\times 10^{-6}$ $(5.11\times 10^{-6}, 6.08\times 10^{-6})$ for the quadratic term.
Given these estimates of the coefficients, the number of spikes increased with the largest depths.
Moreover, as shown in Figure~\ref{fig:calcium_coeff}, the response of neurons is heterogeneous across the cre-lines and, coherent with the results of~\citet{vries2020}, we obtained that the mean response is lower for the VISam, VISpm and VISrl areas.

\subsection{Betting data}
\label{subsec:application2}
Poisson regression models have been widely adopted in sports analytics, where the response variable is the match score or total number of points.
Modelling match scores in association football has recently gained considerable interest owing to the popularity of the betting market, and several modelling approaches have been proposed. Classical methods use only the information of the teams       playing \citep{dixon1997, petretta2021}, while other authors have explored the possibility of introducing additional information, such as historical data and bookmakers’ odds~\citep{egidi2018,Groll2019}.
In general, to describe the number of goals scored by each team in a match, the Poisson distribution is considered a valid assumption \citep{maher1982, lee1997}.
Herein, we considered data of match scores from the Italian Serie A 2020-2021 season, which are publicly available at \url{http://www.football-data.co.uk}. We considered the number of goals as the variable of interest, and, as covariates, we included the fixed effects of the team, several betting odds (for different betting types and bookmakers), and an indicator of whether the team is playing home. The resulting data set has 760 observations on 102 variables.

\begin{figure}
	\centering
	\includegraphics[width = \linewidth]{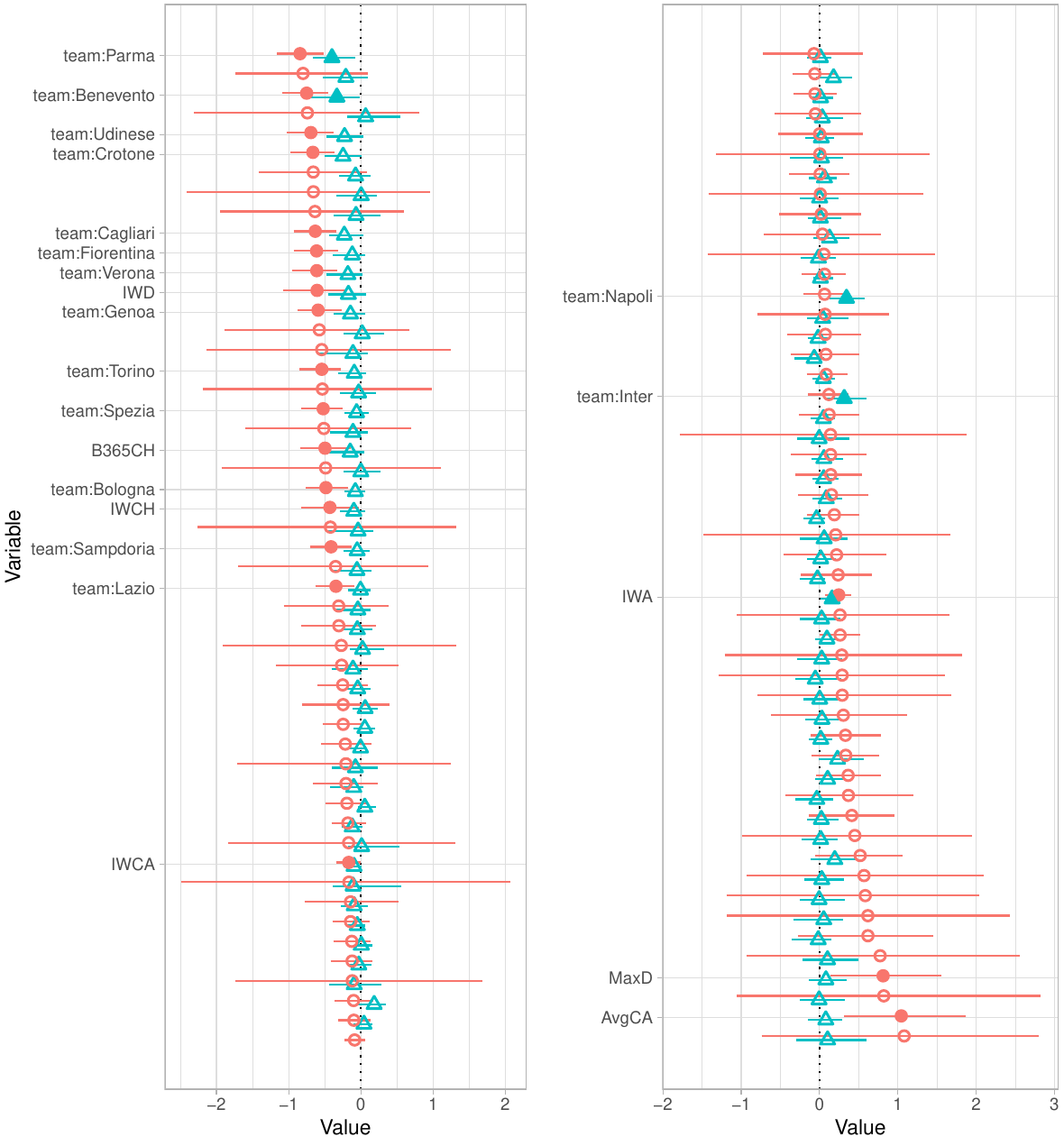}
	\caption{Betting data: posterior mean and 90\% credible interval of the coefficients, obtained using a Gaussian (dots) and horseshoe (triangles) prior. Filled points indicate that the credible interval does not contain zero. Only the names of the non-zero coefficients are shown.}
	\label{fig:football_coef}
\end{figure}

Following the same reasoning as in the previous section, we used the proposed Metropolis-Hastings algorithm. Moreover, we considered both a Gaussian prior centered at zero and a horseshoe prior distribution on the parameters, to compare the results and analyze the variables the two priors select. The results are depicted in Figure~\ref{fig:football_coef}. For each explanatory variable, the posterior mean and credible interval were obtained using the two priors. The filled symbols indicate that the credible interval does not contain zero, showing that the shrinkage induced by the horseshoe prior selects only a few variables compared to the informative Gaussian prior. Moreover, the horseshoe prior induces a significant reduction of the amplitude of all credible intervals.

A more formal comparison between the two models is obtained using the conditional predictive ordinate (CPO) statistics~\citep{geisser1993,gelfand1992,gelfanddey1994}, which is defined for $i=1,\dots,n$, as 
$\text{CPO}_i = p(y_i \mid y_{-i})$,
where $y_{-i}$ is the vector of observed data omitting the $i$-th value. 
Figure~\ref{fig:football_cpo} shows the boxplots of the resulting CPO statistics for the two models. The graph does not highlight any fundamental difference in the predictive capacity of the two models, implying that the horseshoe prior allows to obtain a more parsimonious model with a similar fit. This is also confirmed by the logarithm of the pseudo-marginal likelihood , which is commonly used as a summary of CPO's~\citep{ibrahim203}. It is equal to -1172.055 and to -1167.521 for the models based on the Gaussian prior and the horseshoe, respectively.

\begin{figure}
	\centering
	\includegraphics[width = .5\linewidth]{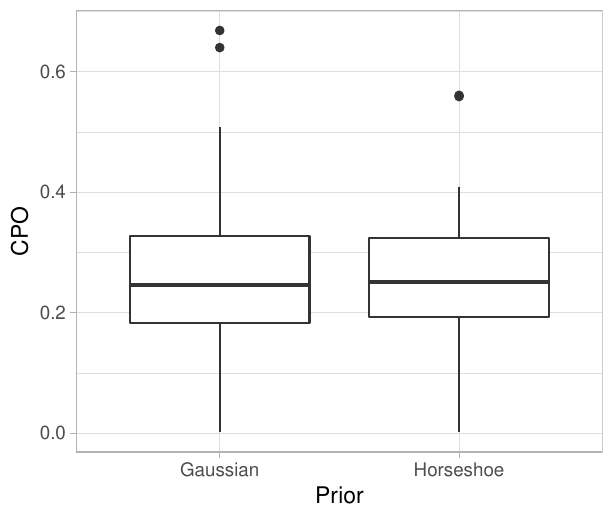}
	\caption{Betting data: distribution of the CPO under the two prior distributions.}
	\label{fig:football_cpo}
\end{figure}

\section{Discussion}

Motivated by the lack of specific computational tools for efficient sampling from the posterior distribution of regression parameters in Poisson log-linear models, we introduced an approximate posterior distribution used as the building block for the Metropolis-Hastings and importance sampling algorithms. 

The introduced proposal distribution is based on two levels of approximation of the target density. The first level exploits the well-known convergence of the negative binomial likelihood to the Poisson likelihood; the second level introduces a Gaussian approximation of the data augmentation scheme of \citet{polson_scott_2013} to sample from this negative binomial model.
The accuracy of the overall approximation can be tuned by acting on the first of these two levels only: thanks to the availability of a simple expression of the relative error between the Poisson and negative binomial distribution, it is possible to fix an upper bound for the discrepancy between the two likelihoods. Although we do not have a handle on the second approximation level, the successful empirical results that we showed are reassuring. 

The resulting density is particularly convenient for use as proposal distribution in Monte Carlo algorithms for two reasons: the possibility to tune its closeness to the target density through a single parameter, and the extremely simple form, which is multivariate normal.

The performances of the proposed solutions, in terms of mixing and computation time, were comparable or superior to those of the efficient Stan implementation of HMC in all scenarios considered and particularly when a hierarchical prior is assumed. 
The ease of application of our methods is further enhanced by their availability via the R package \texttt{bpr}, which obtains the posterior distribution of several quantities of interest without the need for coding and with minimal tuning.

\bigskip
\section*{Supplementary Materials}

\begin{description}	
	\item[Appendix A:] Derivation of the data augmentation scheme for the negative binomial model.
	\item[Appendix B:] Simulation results for $p$ up to 50 covariates.
\end{description}

\bibliographystyle{agsm} \bibliography{biblio.bib}

@article{polson_scott_2013,
	author = {Nicholas G. Polson and James G. Scott and Jesse Windle},
	title = {Bayesian Inference for Logistic Models Using {P\'olya}-Gamma Latent Variables},
	journal = {Journal of the American Statistical Association},
	volume = {108},
	number = {504},
	pages = {1339-1349},
	year  = {2013},
	publisher = {Taylor & Francis}
}

@article{albert_chib_1993,
	author = { James H. Albert and Siddhartha Chib },
	title = {Bayesian Analysis of Binary and Polychotomous Response Data},
	journal = {Journal of the American Statistical Association},
	volume = {88},
	number = {422},
	pages = {669-679},
	year  = {1993},
	publisher = {Taylor & Francis}
}

@article{Teerapabolarn2012,
	author = {K. Teerapabolarn },
	title = {The Least Upper Bound on the {Poisson}-Negative Binomial Relative Error},
	journal = {Communications in Statistics - Theory and Methods},
	volume = {41},
	number = {10},
	pages = {1833-1838},
	year  = {2012},
	publisher = {Taylor & Francis}
}

@article{frome1985,
    author = {Frome, Edward L. and Checkoway, Harvey},
    title = {{Use of Poisson regression models in estimating incidence rates and ratios}},
    journal = {American Journal of Epidemiology},
    volume = {121},
    number = {2},
    pages = {309-323},
    year = {1985},
    month = {02}
}

@inproceedings{chan2009,  
    author={A. B. {Chan} and N. {Vasconcelos}},  
    booktitle={2009 IEEE 12th International Conference on Computer Vision},   
    title={Bayesian {Poisson} regression for crowd counting},   
    year={2009},  
    volume={},  
    number={},  
    pages={545-551}
}

@article{fruhwirth2006,
 author = {Sylvia Fr\"uhwirth-Schnatter and Helga Wagner},
 journal = {Biometrika},
 number = {4},
 pages = {827-841},
 publisher = {[Oxford University Press, Biometrika Trust]},
 title = {Auxiliary Mixture Sampling for Parameter-Driven Models of Time Series of Counts with Applications to State Space Modelling},
 volume = {93},
 year = {2006}
}

@article{fruhwirth2009,
     author = {Sylvia Fr\"uhwirth-Schnatter and Rudolf Fr\"uhwirth and Leonhard Held and Håvard Rue },
     journal = {Statistics and Computing},
     number = {479},
     publisher = {[Oxford University Press, Biometrika Trust]},
     title = {{Improved auxiliary mixture sampling for hierarchical models of non-Gaussian data}},
     volume = {19},
     year = {2009}
}

@article{karlis2005,
    author = "Karlis, D. and Meligkotsidou, L.",
    journal = {Statistics and Computing},
    pages = "255-265",
    title = {{Multivariate Poisson regression with covariance structure}},
    volume = "15",
    year = {2005}
}

@article{bradley2018,
    author = "Bradley, Jonathan R. and Holan, Scott H. and Wikle, Christopher K.",
    doi = "10.1214/17-BA1069",
    journal = "Bayesian Analysis",
    month = "03",
    number = "1",
    pages = "253-310",
    publisher = "International Society for Bayesian Analysis",
    title = "Computationally Efficient Multivariate Spatio-Temporal Models for High-Dimensional Count-Valued Data (with Discussion)",
    volume = "13",
    year = "2018"
}

@article{park2008bayesian,
  title={The {B}ayesian lasso},
  author={Park, Trevor and Casella, George},
  journal={Journal of the American Statistical Association},
  volume={103},
  number={482},
  pages={681--686},
  year={2008},
  publisher={Taylor \& Francis}
}

@article{carvalho2010horseshoe,
  title={The horseshoe estimator for sparse signals},
  author={Carvalho, Carlos M and Polson, Nicholas G and Scott, James G},
  journal={Biometrika},
  volume={97},
  number={2},
  pages={465--480},
  year={2010},
  publisher={Oxford University Press}
}

@ARTICLE{makalic2016horseshoesampler,
  author={E. Makalic and D. F. Schmidt},
  journal={IEEE Signal Processing Letters}, 
  title={A Simple Sampler for the Horseshoe Estimator}, 
  year={2016},
  volume={23},
  number={1},
  pages={179-182}}

@article{vanderpas2017,
    author = {van der Pas, Stéphanie and Szabó, Botond and van der Vaart, Aad},
    title = {{Adaptive posterior contraction rates for the horseshoe}},
    volume = {11},
    journal = {Electronic Journal of Statistics},
    number = {2},
    publisher = {Institute of Mathematical Statistics and Bernoulli Society},
    pages = {3196 -- 3225},
    year = {2017}
}

@article{nelder1972glm,
 author = {J. A. Nelder and R. W. M. Wedderburn},
 journal = {Journal of the Royal Statistical Society. Series A (General)},
 number = {3},
 pages = {370--384},
 publisher = {[Royal Statistical Society, Wiley]},
 title = {Generalized Linear Models},
 volume = {135},
 year = {1972}
}

@article{Frome1983,
 author = {E. L. Frome},
 journal = {Biometrics},
 number = {3},
 pages = {665--674},
 publisher = {[Wiley, International Biometric Society]},
 title = {The Analysis of Rates Using {Poisson} Regression Models},
 volume = {39},
 year = {1983}
}

@article{lambert1992,
    author = { Diane Lambert },
    title = {Zero-Inflated {Poisson} Regression, With an Application to Defects in Manufacturing},
    journal = {Technometrics},
    volume = {34},
    number = {1},
    pages = {1-14},
    year  = {1992},
    publisher = {Taylor & Francis}
}

@article{Sarath1990,
    author = { Sarath C. Joshua  and  Nicholas J. Garber },
    title = {Estimating truck accident rate and involvements using linear and {Poisson} regression models},
    journal = {Transportation Planning and Technology},
    volume = {15},
    number = {1},
    pages = {41-58},
    year  = {1990},
    publisher = {Routledge}
}

@article{Hutchinson2005,
    author = {Hutchinson, M. Katherine and Holtman, Matthew C.},
    title = {Analysis of count data using {Poisson} regression},
    journal = {Research in Nursing \& Health},
    volume = {28},
    number = {5},
    pages = {408-418},
    year = {2005}
}

@article{Miaou1994,
    title = {The relationship between truck accidents and geometric design of road sections: {Poisson} versus negative binomial regressions},
    author = {Shaw-Pin Miaou},
    journal = {Accident Analysis \& Prevention},
    volume = {26},
    number = {4},
    pages = {471-482},
    year = {1994}
}

@article{Piironen2017,
    author = {Juho Piironen and Aki Vehtari},
    title = {{Sparsity information and regularization in the horseshoe and other shrinkage priors}},
    volume = {11},
    journal = {Electronic Journal of Statistics},
    number = {2},
    publisher = {Institute of Mathematical Statistics and Bernoulli Society},
    pages = {5018 -- 5051},
    year = {2017}
}

@book{robert2010,
    author = {Christian Robert and George Casella},
    title = {{Introducing Monte Carlo methods with R}},
    publisher = {Springer},
    year = {2010}
}

@misc{allen,
	author = {{Allen Institute for Brain Science}},
	howpublished = {http://observatory.brain-map.org/visualcoding},
	title = {Allen Brain Observatory},
	year = {2016}}

@article{jewell2019,
	author = {Jewell, Sean W. and Hocking, Toby Dylan and Fearnhead, Paul and Witten, Daniela M.},
	journal = {Biostatistics},
	month = {02},
	number = {4},
	pages = {709-726},
	title = {Fast nonconvex deconvolution of calcium imaging data},
	volume = {21},
	year = {2019}
}

@article{vries2020,
    author = {de Vries, Saskia and Lecoq, Jerome and Buice, Michael and Groblewski, Peter and Ocker, Gabriel and Oliver, Michael and Feng, David and Cain, Nicholas and Ledochowitsch, Peter and Millman, Daniel and Roll, Kate and Garrett, Marina and Keenan, Tom and Kuan, Chihchau and Mihalas, Stefan and Olsen, Shawn and Thompson, Carol and Wakeman, Wayne and Waters, Jack and Koch, Christof},
    year = {2020},
    pages = {138-151},
    title = {A large-scale standardized physiological survey reveals functional organization of the mouse visual cortex},
    journal = {Nature neuroscience},
    volume = {23},
    number = {1}
}

@incollection{paninski2007,
	title = {Statistical models for neural encoding, decoding, and optimal stimulus design},
	editor = {Paul Cisek and Trevor Drew and John F. Kalaska},
	series = {Progress in Brain Research},
	publisher = {Elsevier},
	volume = {165},
	pages = {493-507},
	year = {2007},
	booktitle = {Computational Neuroscience: Theoretical Insights into Brain Function},
	author = {Liam Paninski and Jonathan Pillow and Jeremy Lewi}
}

@article{maher1982,
    author = {Maher, M. J.},
    year = {1982},
    pages = {109–118},
    title = {Modelling association football scores},
    journal = {Statistica Neerlandica},
    volume = {36},
    number = {3}
}

@article{lee1997,
    author = { Alan J. Lee },
    title = {Modeling Scores in the Premier League: Is {Manchester United} Really the Best?},
    journal = {Chance},
    volume = {10},
    number = {1},
    pages = {15-19},
    year  = {1997}
}

@article{egidi2018,
    author = {Leonardo Egidi and Francesco Pauli and Nicola  Torelli},
    title ={Combining historical data and bookmakers’ odds in modelling football scores},
    journal = {Statistical Modelling},
    volume = {18},
    number = {5-6},
    pages = {436-459},
    year = {2018}
}

@article{petretta2021,
      title={Mar-Co: a new dependence structure to model match outcomes in football}, 
      author={Marco Petretta and Lorenzo Schiavon and Jacopo Diquigiovanni},
      year={2021},
      eprint={2103.07272},
      journal={arXiv:2103.07272},
      primaryClass={stat.ME}
}

@article{dixon1997,
    author = {Dixon, Mark J. and Coles, Stuart G.},
    title = {Modelling Association Football Scores and Inefficiencies in the Football Betting Market},
    journal = {Journal of the Royal Statistical Society: Series C (Applied Statistics)},
    volume = {46},
    number = {2},
    pages = {265-280},
    year = {1997}
}

@article{Groll2019,
author = {Andreas Groll and Cristophe Ley and Gunther Schauberger and Hans Van Eetvelde},
title = {A hybrid random forest to predict soccer matches in international tournaments},
journal = {Journal of Quantitative Analysis in Sports},
number = {4},
volume = {15},
year = {2019},
pages = {271--287}
}

@inproceedings{gelfand1992,
    author = {Gelfand, A. and Dey, Dipak and Chang, Hong},
    title = {Model determination using predictive distributions with implementation via sampling-based-methods (with Discussion)},
    booktitle = {Bayesian Statistics 4},
    year = {1992},
    publisher = {University Press}
}

@article{gelfanddey1994,
 author = {A. E. Gelfand and D. K. Dey},
 journal = {Journal of the Royal Statistical Society. Series B (Methodological)},
 number = {3},
 pages = {501--514},
 publisher = {[Royal Statistical Society, Wiley]},
 title = {Bayesian model choice: asymptotics and exact calculations},
 volume = {56},
 year = {1994}
}

@inbook{ibrahim203,
author = {Ibrahim, Joseph G. and Chen, Ming-Hui and Sinha, Debajyoti},
publisher = {American Cancer Society},
isbn = {9781118445112},
title = {Bayesian Survival Analysis},
booktitle = {Wiley StatsRef: Statistics Reference Online},
year = {2014}
}

@book{geisser1993,
    author = {Geisser, S.},
    year= {1993},
    title={Predictive Inference},
    publisher = {Chapman and Hall/CRC}
}

@article{hastings1970monte,
    author = {W. K. Hastings},
    journal = {Biometrika},
    number = {1},
    pages = {97--109},
    title = {Monte {Carlo} Sampling Methods Using {Markov} Chains and Their Applications},
    volume = {57},
    year = {1970}
}

@article{neal2011mcmc,
  title={{MCMC} using {Hamiltonian} dynamics},
  author={Neal, Radford M},
  journal={Handbook of {M}arkov chain {M}onte {C}arlo},
  volume={2},
  number={11},
  pages={2},
  year={2011}
}

@Misc{stan,
    title = {Stan Modeling Language Users Guide and Reference Manual},
    author = {{Stan Development Team}},
    year = {2021},
    howpublished = {\textsc{url:} \texttt{http://mc-stan.org/}}
}

@article{RCPP,
   author = {Dirk Eddelbuettel and Romain Francois},
   title = {Rcpp: Seamless {R} and {C}++ Integration},
   journal = {Journal of Statistical Software, Articles},
   volume = {40},
   number = {8},
   year = {2011},
   issn = {1548-7660},
   pages = {1--18}
}

@article{W,
   author = {Lambert, J. H.},
   title = {Observations variae in Mathesin Puram},
   journal = {Acta Helvitica, physico-mathematico-anatomico-botanico-medica},
   volume = {3},
   year = {1758},
   pages = {128-168}
}

@misc{bpr,
   author = {Laura D'Angelo},
   title = {{bpr: Bayesian Poisson regression}},
   year = {2021},
   howpublished = {\textsc{url:} https://CRAN.R-project.org/package=bpr}
}

@article{arridge2018,
   author = {Simon R. Arridge and Kazufumi Ito and Bangti Jin and Chen Zhang},
   title = {{Variational Gaussian approximation for Poisson data}},
   journal = {Inverse Problems},
   volume = {34},
   number = {2},
   year = {2018},
   pages = {1--29}
}
 
 \clearpage
\appendix
\section*{Appendix A: Derivation of the data augmentation scheme}
\label{sec:appA}
In this section, we derive the data augmentation scheme based on P\'olya-gamma random variables for the negative binomial model in Equation (2) of the paper, where we make explicit the mean parameter $\lambda_i = e^{x_i^T\beta}$
\begin{align*}
\tilde{f}_{r_i}(y_i \mid \beta) &=
\binom{r_i + y_i -1}{r_i-1} \left( \frac{r_i}{r_i + e^{x_i^T\beta}}\right)^{r_i} \left(\frac{e^{x_i^T\beta}}{r_i+ e^{x_i^T\beta}}\right)^{y_i} \\
&= \binom{r_i + y_i -1}{r_i-1} \, r_i^{r_i} \, \frac{(e^{x_i^T\beta})^{y_i}}{(r + e^{x_i^T\beta})^{r_i+y_i}} \\
&= \binom{r_i + y_i -1}{r_i-1} \, r_i^{r_i} \, \frac{r_i^{y_i}}{r_i^{r_i+y_i}} \, \frac{(e^{x_i^T\beta - \log r_i})^{y_i}}{(1 + e^{x_i^T\beta - \log r_i})^{r_i+y_i}} \\
&= \binom{r_i + y_i -1}{r_i-1}  \frac{(e^{x_i^T\beta - \log r_i})^{y_i}}{(1 + e^{x_i^T\beta - \log r_i})^{r_i+y_i}}.
\end{align*}
From this form of the likelihood, it is immediate to obtain Equation (3) by simply adjusting the parameters of the original data augmentation of \citet{polson_scott_2013}.

\bigskip

\section*{Appendix B: Simulation results for $p$ up to 50 covariates}
\label{sec:appB}
In this section, we compare the performances of the proposed algorithms for a number of covariates $p=5, 10, 20, 50$. Specifically, similarly to the main text, we analyze the logarithm of the time per independent sample for the proposed Metropolis-Hastings algorithm and importance sampler, and the Stan implementation of the HMC \citep{stan}.
For each combination we only plot the median time per independent sample (instead of a boxplot, as in the paper), for graphical reasons and ease of interpretation.

\begin{figure}
	\centering
	\includegraphics[width = \linewidth]{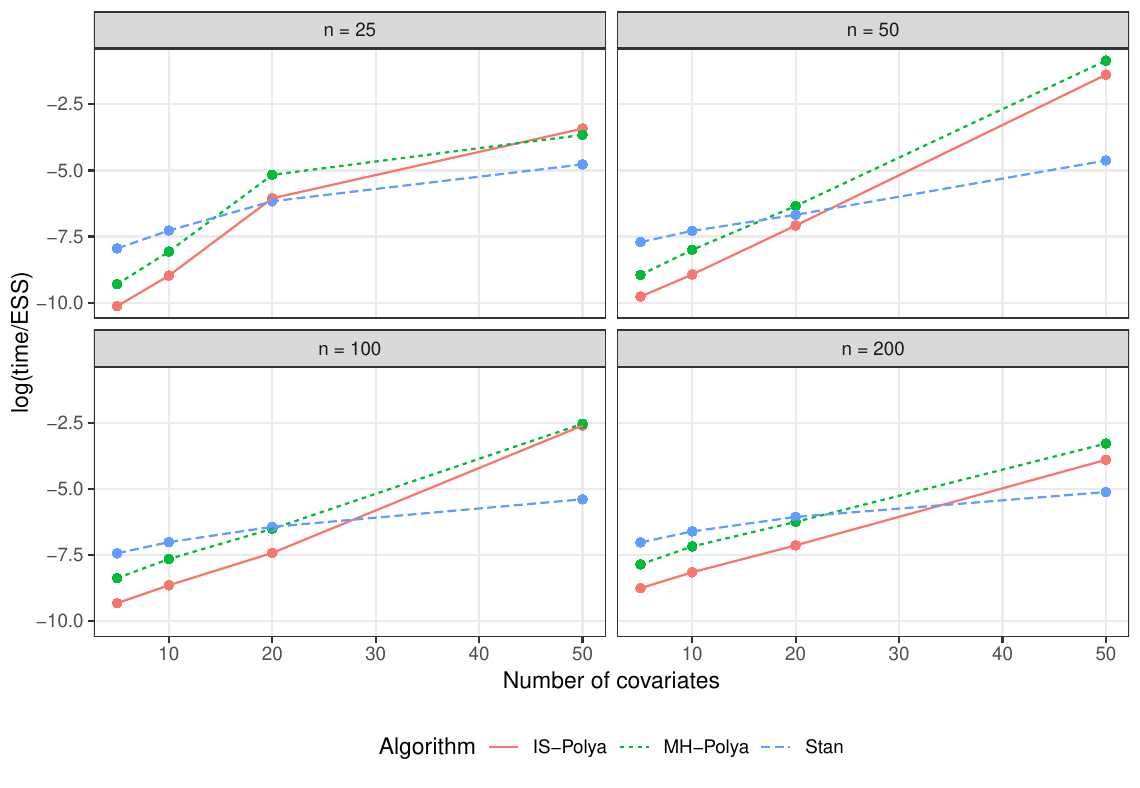}
	\caption{Time per independent sample (in logarithmic scale) for the three algorithms. For each combination of $n$ and $p$ the point is the median of the (log) time (in seconds) over the effective sample size using a Gaussian prior, over 50 replications.}
\end{figure}

\begin{figure}
	\centering
	\includegraphics[width = \linewidth]{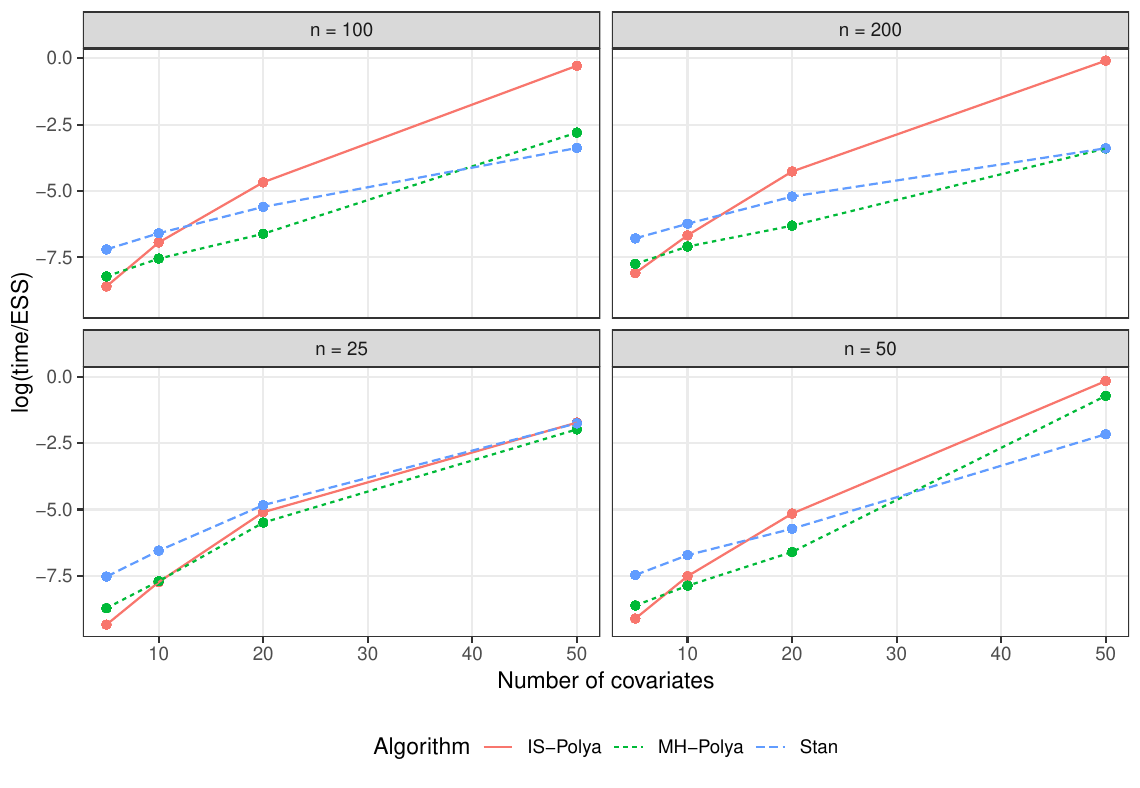}
	\caption{Time per independent sample (in logarithmic scale) for the three algorithms. For each combination of $n$ and $p$ the point is the median of the (log) time (in seconds) over the effective sample size using the horseshoe prior, over 50 replications.}
\end{figure}

\end{document}